\documentclass{article}
\usepackage{amsmath}

\input{tcilatex}
\begin{document}

\title{Time-Fractional KdV Equation for the plasma in auroral zone using
Variational Methods}
\date{}
\author{El-Said A. El-Wakil, Essam M. Abulwafa, \and Emad K. Elshewy and
Aber A. Mahmoud \\
Theoretical Physics Research Group, Physics Department, \\
Faculty of Science, Mansoura University, Mansoura 35516, Egypt}
\maketitle

\begin{abstract}
The reductive perturbation method has been employed to derive the
Korteweg-de Vries (KdV) equation for small but finite amplitude
electrostatic waves. The Lagrangian of the time fractional KdV equation is
used in similar form to the Lagrangian of the regular KdV equation. The
variation of the functional of this Lagrangian leads to the Euler-Lagrange
equation that leads to the time fractional KdV equation. The
Riemann-Liouvulle definition of the fractional derivative is used to
describe the time fractional operator in the fractional KdV equation. The
variational-iteration method given by He is used to solve the derived time
fractional KdV equation. The calculations of the solution with initial
condition $A_{0}\sec h(cx)^{2}$ are carried out. Numerical studies have been
made using plasma parameters close to those values corresponding to the
dayside auroral zone. The effects of the time fractional parameter on the
electrostatic solitary structures are presented.

\begin{description}
\item Keywords: Euler-Lagrange equation, Riemann-Liouvulle fractional
derivative, fractional KdV equation, He's variational-iteration method.

\item PACS: 05.45.Df, 05.30.Pr
\end{description}
\end{abstract}

\section{Introduction}

Because most classical processes observed in the physical world are
nonconservative, it is important to be able to apply the power of
variational methods to such cases. A method used a Lagrangian that leads to
an Euler-Lagrange equation that is, in some sense, equivalent to the desired
equation of motion. Hamilton's equations are derived from the Lagrangian and
are equivalent to the Euler-Lagrange equation. If a Lagrangian is
constructed using noninteger-order derivatives, then the resulting equation
of motion can be nonconservative. It was shown that such fractional
derivatives in the Lagrangian describe nonconservative forces [1, 2].
Further study of the fractional Euler-Lagrange can be found in the work of
Agrawal [3, 4], Baleanu and coworkers [5, 6] and Tarasov and Zaslavsky [7,
8]. During the last decades, Fractional Calculus has been applied to almost
every field of science, engineering and mathematics. Some of the areas where
Fractional Calculus has been applied include viscoelasticity and rheology,
electrical engineering, electrochemistry, biology, biophysics and
bioengineering, signal and image processing, mechanics, mechatronics,
physics, and control theory [9].

On the other hand, electron acoustic waves (EAWs) propagation in plasmas has
received a great deal attention because of its vital role in understanding
different types of collective processes in laboratory devices [10, 11] as
well as in space environments [12, 13]. It has been argued that when the hot
to cold electron temperature ratio is greater than 10, the electron-acoustic
mode may be the principal mode of the plasma in which the restoring force
comes from the pressure of the hot electrons, while the inertia comes from
the mass of the cold electron component [14]. The ions play the role of a
neutralizing background, i.e., the ion dynamics does not influence the EAWs
because its frequency is much larger than the ion plasma frequency. Several
theoretical attempts have been made to explain nonlinear EAWs in plasma
systems [15-17].

To the author's knowledge, the problem of time fractional KdV equation in
collisionless plasma has not been addressed in the literature before. So,
our motive here is to study the effects of time fractional parameter on the
electrostatic structures for a system of collisionless plasma consisting of
a cold electron fluid and non-thermal hot electrons obeying a non-thermal
distribution and stationary ions. Our choice of non-thermal distribution of
electrons is prompted by its convenience rather than as precise fitting of
the observations. We expect that the inclusion of the non-thermal electrons,
time fractional parameter will change the properties as well as the regime
of existence of solitons.

Several methods have been used to solve fractional differential equations
such as: the Laplace transform method, the Fourier transform method, the
iteration method and the operational method [18, 19]. Recently, there are
some papers deal with the existence and multiplicity of solution of
nonlinear fractional differential equation by the use of techniques of
nonlinear analysis [20-23]. In this paper, the resultant fractional KdV
equation will be solved using a variational-iteration method (VIM) firstly
used by He [24, 25].

This paper is organized as follows: Section 2 is devoted to describe the
formulation of the time-fractional KdV (FKdV) equation using the variational
Euler-Lagrange method. In section 3, the resultant time-FKdV equation is
solved approximately using VIM. Section 4 contains the results of
calculations and discussion of these results.

\section{Basic Equations and Time-fractional KdV equation}

We consider a homogeneous system of unmagnetized collisionless plasma
consisting of cold electrons fluid, non-thermal hot electrons obeying a
non-thermal distribution and stationary ions. Such system is governed by the
following normalized equations in one dimension [26]:

\begin{equation}
\frac{\partial }{\partial t}u_{c}(x,~t)+\frac{\partial }{\partial x}%
[u_{c}(x,~t)~n_{c}(x,~t)]=0\text{,}  \tag{1a}
\end{equation}

\begin{equation}
\frac{\partial }{\partial t}u_{c}(x,~t)+u_{c}(x,~t)\frac{\partial }{\partial
t}u_{c}(x,~t)-\gamma \phi (x,~t)=0\text{,}  \tag{1b}
\end{equation}

with Poisson's equation

\begin{equation}
\frac{\partial ^{2}}{\partial x^{2}}\phi (x,~t)-\frac{1}{\gamma }%
n_{c}(x,~t)-n_{h}(x,~t)+(1+\frac{1}{\gamma })=0\text{.}  \tag{1c}
\end{equation}

The non-thermal hot electrons density $n_{h}(x,~t)$ is given by:

\begin{equation}
n_{h}(x,~t)=[1-\beta ~\phi (x,~t)+\beta ~\phi (x,~t)^{2}]~\exp [\phi (x,~t)]%
\text{, }\beta =4\delta /(1+3\delta )\text{.}  \tag{2}
\end{equation}

In these equations, $n_{c}(x,~t)$[$n_{h}(x,~t)$] is the cold (non-thermal
hot) electrons density normalized by equilibrium value $n_{c0}$[$n_{h0}$], $%
u_{c}(x,~t)$ is the cold electrons fluid velocity normalized by hot electron
acoustic speed $C_{e}=\sqrt{\frac{k_{B}T_{h}}{\gamma m_{e}}}$, $\phi (x,~t)$
is the electric potential normalized by $\frac{k_{B}T_{h}}{e}$, $\gamma =%
\frac{n_{h0}}{n_{c0}}$ is the hot to cold electron equilibrium densities
ratio, $m_{e}$ is the electron mass, $\delta $ is a parameter which
determines the population of energetic non-thermal hot electrons, $e$ is the
electron charge, $x$ is the space co-ordinate and $t$ is the time variable.
The distance is normalized to the hot electron Debye length $\lambda _{Dh}$,
the time is normalized by the inverse of the cold electron plasma frequency $%
\omega _{ce}^{-1}$ and $k_{B}$ is the Boltzmann's constant. Equations (1a)
and (1b) represent the inertia of cold electron and (1c) is the Poisson's
equation needed to make the self consistent. The hot electrons are described
by non-thermal distribution given by (2).

According to the general method of reductive perturbation theory, we
introduce the slow stretched co-ordinates [27]:

\begin{equation}
\tau =\epsilon ^{2/3}t\text{ and }\xi =\epsilon ^{1/2}(x-\lambda t)\text{,} 
\tag{3}
\end{equation}%
where $\epsilon $ is a small dimensionless expansion parameter and $\lambda $
is the wave speed normalized by $C_{e}$. All physical quantities appearing
in (1) are expanded as power series in $\epsilon $ about their equilibrium
values as:

\begin{equation}
n_{c}(\xi ,~\tau )=1+\epsilon n_{1}(\xi ,~\tau )+\epsilon ^{2}n_{2}(\xi
,~\tau )+\epsilon ^{3}n_{3}(\xi ,~\tau )+...\text{,}  \tag{4a}
\end{equation}

\begin{equation}
u_{c}(\xi ,~\tau )=\epsilon u_{1}(\xi ,~\tau )+\epsilon ^{2}u_{2}(\xi ,~\tau
)+\epsilon ^{3}u_{3}(\xi ,~\tau )+...\text{,}  \tag{4b}
\end{equation}

\begin{equation}
\phi (\xi ,~\tau )=\epsilon \phi _{1}(\xi ,~\tau )+\epsilon _{{}}^{2}\phi
_{2}(\xi ,~\tau )+\epsilon _{{}}^{3}\phi _{3}(\xi ,~\tau )+...\text{.} 
\tag{4c}
\end{equation}

We impose the boundary conditions as $\xi \rightarrow \infty $, $%
n_{c}=n_{h}=1$, $u_{c}=0$ and $\phi =0$.

Substituting (3) and (4) into (1) and equating coefficients of like powers
of $\epsilon $, the lowest-order equations in $\epsilon $ lead to the
following results:

\begin{equation}
n_{1}(\xi ,\tau )=\frac{\gamma }{\lambda ^{2}}\phi _{1}(\xi ,\tau )\text{
and }u_{1}(\xi ,\tau )=\frac{\gamma }{\lambda }\phi _{1}(\xi ,\tau )\text{.}
\tag{5}
\end{equation}

Poisson's equation gives the linear dispersion relation

\begin{equation}
\lambda =\sqrt{\frac{1}{1-\beta }}=\sqrt{\frac{1+3\delta }{1-\delta }}\text{.%
}  \tag{6}
\end{equation}

Considering the coefficients of $O(\epsilon ^{2})$ and eliminating the
second order-perturbed quantities $n_{2}$, $u_{2}$ and $\phi _{2}$ lead to
the following KdV equation for the first-order perturbed potential:

\begin{equation}
\frac{\partial }{\partial \tau }\phi _{1}(\xi ,~\tau )+A~\phi _{1}(\xi
,~\tau )\frac{\partial }{\partial \xi }\phi _{1}(\xi ,~\tau )+B~\frac{%
\partial ^{3}}{\partial \xi ^{3}}\phi _{1}(\xi ,~\tau )=0\text{,}  \tag{7a}
\end{equation}%
where

\begin{equation}
A=-\frac{3\gamma +\lambda ^{4}}{2\lambda }\text{, }B=\frac{\lambda ^{3}}{2}%
\text{,}  \tag{7b}
\end{equation}
$\phi _{1}(\xi ,~\tau )$ is a field variable, $\xi $ is a space coordinate
in the propagation direction of the field and $\tau \in T$($=[0,T_{0}]$) is
the time. The resultant KdV equation (7a) can be converted into
time-fractional KdV equation as follows:

Using a potential function $v(\xi ,~\tau )$ where $\phi _{1}(\xi ,~\tau
)=v_{\xi }(\xi ,~\tau )$ gives the potential equation of the regular KdV
equation (1) in the form

\begin{equation}
v_{\xi \tau }(\xi ,~\tau )+A~v_{\xi }(\xi ,~\tau )v_{\xi \xi }(\xi ,~\tau
)+B~v_{\xi \xi \xi \xi }(\xi ,~\tau )=0\text{,}  \tag{8}
\end{equation}%
where the subscripts denote the partial differentiation of the function with
respect to the parameter. The Lagrangian of this regular KdV equation (7)
can be defined using the semi-inverse method [28, 29] as follows.

The functional of the potential equation (8) can be represented by

\begin{equation}
J(v)=\dint\limits_{R}d\xi \dint\limits_{T}d\tau \{v(\xi ,\tau )[c_{1}v_{\xi
\tau }(\xi ,\tau )+c_{2}Av_{\xi }(\xi ,\tau )v_{\xi \xi }(\xi ,\tau
)+c_{3}Bv_{\xi \xi \xi \xi }(\xi ,\tau )]\}\text{,}  \tag{9}
\end{equation}%
where $c_{1}$, $c_{2}$ and $c_{3}$ are constants to be determined.
Integrating by parts and taking $v_{\tau }|_{R}=v_{\xi }|_{R}=v_{\xi
}|_{T}=0 $ lead to

\begin{equation}
J(v)=\dint\limits_{R}d\xi \dint\limits_{T}d\tau \{v(\xi ,\tau )[-c_{1}v_{\xi
}(\xi ,\tau )v_{\tau }(\xi ,\tau )-\frac{1}{2}c_{2}Av_{\xi }^{3}(\xi ,\tau
)+c_{3}Bv_{\xi \xi }^{2}(\xi ,\tau )]\}\text{.}  \tag{10}
\end{equation}

The unknown constants $c_{i}$ ($i=$ $1$, $2$, $3$) can be determined by
taking the variation of the functional (10) to make it optimal. Taking the
variation of this functional, integrating each term by parts and make the
variation optimum give the following relation

\begin{equation}
2c_{1}v_{\xi \tau }(\xi ,\tau )+3c_{2}Av_{\xi }(\xi ,\tau )v_{\xi \xi }(\xi
,\tau )+2c_{3}Bv_{\xi \xi \xi \xi }(\xi ,\tau )=0\text{.}  \tag{11}
\end{equation}

As this equation must be equal to equation (8), the unknown constants are
given as

\begin{equation}
c_{1}=1/2\text{, }c_{2}=1/3\text{ and }c_{3}=1/2\text{.}  \tag{12}
\end{equation}

Therefore, the functional given by (10) gives the Lagrangian of the regular
KdV equation as

\begin{equation}
L(v_{\tau },~v_{\xi },v_{\xi \xi })=-\frac{1}{2}v_{\xi }(\xi ,\tau )v_{\tau
}(\xi ,\tau )-\frac{1}{6}Av_{\xi }^{3}(\xi ,\tau )+\frac{1}{2}Bv_{\xi \xi
}^{2}(\xi ,\tau )\text{.}  \tag{13}
\end{equation}

Similar to this form, the Lagrangian of the time-fractional version of the
KdV equation can be written in the form

\begin{eqnarray}
F(_{0}D_{\tau }^{\alpha }v,~v_{\xi },v_{\xi \xi }) &=&-\frac{1}{2}%
[_{0}D_{\tau }^{\alpha }v(\xi ,\tau )]v_{\xi }(\xi ,\tau )-\frac{1}{6}%
Av_{\xi }^{3}(\xi ,\tau )+\frac{1}{2}Bv_{\xi \xi }^{2}(\xi ,\tau )\text{, } 
\notag \\
0 &\leq &\alpha <1\text{,}  \TCItag{14}
\end{eqnarray}%
where the fractional derivative is represented, using the left
Riemann-Liouville fractional derivative definition as [18, 19]

\begin{equation}
_{a}D_{t}^{\alpha }f(t)=\frac{1}{\Gamma (k-\alpha )}\frac{d^{k}}{dt^{k}}%
[\int_{a}^{t}d\tau (t-\tau )^{k-\alpha -1}f(\tau )]\text{, }k-1\leq \alpha
\leq 1\text{, }t\in \lbrack a,b]\text{.}  \tag{15}
\end{equation}

The functional of the time-FKdV equation can be represented in the form

\begin{equation}
J(v)=\dint\limits_{R}d\xi \dint\limits_{T}d\tau F(_{0}D_{\tau }^{\alpha
}v,~v_{\xi },v_{\xi \xi })\text{,}  \tag{16}
\end{equation}%
where the time-fractional Lagrangian $F(_{0}D_{\tau }^{\alpha }v,~v_{\xi
},v_{\xi \xi })$ is defined by (14).

Following Agrawal's method [3, 4], the variation of functional (16) with
respect to $v(\xi ,\tau )$ leads to

\begin{equation}
\delta J(v)=\dint\limits_{R}d\xi \dint\limits_{T}d\tau \{\frac{\partial F}{%
\partial _{0}D_{\tau }^{\alpha }v}\delta _{0}D_{\tau }^{\alpha }v+\frac{%
\partial F}{\partial v_{\xi }}\delta v_{\xi }+\frac{\partial F}{\partial
v_{\xi \xi }}\delta v_{\xi \xi }\}\text{.}  \tag{17}
\end{equation}

The formula for fractional integration by parts reads [3, 18, 19]

\begin{equation}
\int_{a}^{b}dtf(t)_{a}D_{t}^{\alpha
}g(t)=\int_{a}^{t}dtg(t)_{t}D_{b}^{\alpha }f(t)\text{, \ \ \ }f(t)\text{, }%
g(t)\text{ }\in \lbrack a,~b]\text{.}  \tag{18}
\end{equation}%
where $_{t}D_{b}^{\alpha }$, the right Riemann-Liouville fractional
derivative, is defined by [18, 19]

\begin{equation}
_{t}D_{b}^{\alpha }f(t)=\frac{(-1)^{k}}{\Gamma (k-\alpha )}\frac{d^{k}}{%
dt^{k}}[\int_{t}^{b}d\tau (\tau -t)^{k-\alpha -1}f(\tau )]\text{, }k-1\leq
\alpha \leq 1\text{, }t\in \lbrack a,b]\text{.}  \tag{19}
\end{equation}

Integrating the right-hand side of (17) by parts using formula (18) leads to

\begin{equation}
\delta J(v)=\dint\limits_{R}d\xi \dint\limits_{T}d\tau \lbrack _{\tau
}D_{T_{0}}^{\alpha }(\frac{\partial F}{\partial _{0}D_{\tau }^{\alpha }v})-%
\frac{\partial }{\partial \xi }(\frac{\partial F}{\partial v_{\xi }})+\frac{%
\partial ^{2}}{\partial \xi ^{2}}(\frac{\partial F}{\partial v_{\xi \xi }}%
)]\delta v\text{,}  \tag{20}
\end{equation}%
where it is assumed that $\delta v|_{T}=\delta v|_{R}=\delta v_{\xi }|_{R}=0$%
.

Optimizing this variation of the functional $J(v)$, i. e; $\delta J(v)=0$,
gives the Euler-Lagrange equation for the time-FKdV equation in the form

\begin{equation}
_{\tau }D_{T_{0}}^{\alpha }(\frac{\partial F}{\partial _{0}D_{\tau }^{\alpha
}v})-\frac{\partial }{\partial \xi }(\frac{\partial F}{\partial v_{\xi }})+%
\frac{\partial ^{2}}{\partial \xi ^{2}}(\frac{\partial F}{\partial v_{\xi
\xi }})=0\text{.}  \tag{21}
\end{equation}

Substituting the Lagrangian of the time-FKdV equation (14) into this
Euler-Lagrange formula (21) gives

\begin{equation}
-\frac{1}{2}~_{\tau }D_{T_{0}}^{\alpha }v_{\xi }(\xi ,\tau )+\frac{1}{2}%
~_{0}D_{\tau }^{\alpha }v_{\xi }(\xi ,\tau )+Av_{\xi }(\xi ,\tau )v_{\xi \xi
}(\xi ,\tau )+Bv_{\xi \xi \xi \xi }(\xi ,\tau )=0\text{.}  \tag{22}
\end{equation}

Substituting for the potential function, $v_{\xi }(\xi ,\tau )=\phi _{1}(\xi
,\tau )=\Phi (\xi ,\tau )$, gives the time-FKdV equation for the state
function $\Phi (\xi ,\tau )$ in the form

\begin{equation}
\frac{1}{2}[_{0}D_{\tau }^{\alpha }\Phi (\xi ,\tau )-_{\tau
}D_{T_{0}}^{\alpha }\Phi (\xi ,\tau )]+A\Phi (\xi ,\tau )\Phi _{\xi }(\xi
,\tau )+B\Phi _{\xi \xi \xi }(\xi ,\tau )=0\text{,}  \tag{23}
\end{equation}%
where the fractional derivatives $_{0}D_{\tau }^{\alpha }$ and $_{\tau
}D_{T_{0}}^{\alpha }$ are, respectively the left and right Riemann-Liouville
fractional derivatives and are defined by (15) and (19).

The time-FKdV equation represented in (14) can be rewritten by the formula

\begin{equation}
\frac{1}{2}~_{0}^{R}D_{\tau }^{\alpha }\Phi (\xi ,\tau )+A~\Phi (\xi ,\tau
)\Phi _{\xi }(\xi ,\tau )+B~\Phi _{\xi \xi \xi }(\xi ,\tau )=0\text{,} 
\tag{24}
\end{equation}%
where the fractional operator $_{0}^{R}D_{\tau }^{\alpha }$ is called Riesz
fractional derivative and can be represented by [4, 18, 19]

\begin{eqnarray}
~_{0}^{R}D_{t}^{\alpha }f(t) &=&\frac{1}{2}[_{0}D_{t}^{\alpha
}f(t)+~(-1)^{k}{}_{t}D_{T_{0}}^{\alpha }f(t)]  \notag \\
&=&\frac{1}{2}\frac{1}{\Gamma (k-\alpha )}\frac{d^{k}}{dt^{k}}%
[\int_{a}^{t}d\tau |t-\tau |^{k-\alpha -1}f(\tau )]\text{, }  \notag \\
k-1 &\leq &\alpha \leq 1\text{, }t\in \lbrack a,b]\text{.}  \TCItag{25}
\end{eqnarray}

The nonlinear fractional differential equations have been solved using
different techniques [18-23]. In this paper, a variational-iteration method
(VIM) [24, 25] has been used to solve the time-FKdV equation that formulated
using Euler-Lagrange variational technique.

\section{Variational-Iteration Method}

Variational-iteration method (VIM) [24, 25] has been used successfully to
solve different types of integer nonlinear differential equations [30, 31].
Also, VIM is used to solve linear and nonlinear fractional differential
equations [32, 33]. This VIM has been used in this paper to solve the
formulated time-FKdV equation.

A general Lagrange multiplier method is constructed to solve non-linear
problems, which was first proposed to solve problems in quantum mechanics
[24]. The VIM is a modification of this Lagrange multiplier method [25]. The
basic features of the VIM are as follows. The solution of a linear
mathematical problem or the initial (boundary) condition of the nonlinear
problem is used as initial approximation or trail function. A more highly
precise approximation can be obtained using iteration correction functional.
Considering a nonlinear partial differential equation consists of a linear
part $\overset{\symbol{94}}{L}U(x,t)$, nonlinear part $\overset{\symbol{94}}{%
N}U(x,t)$ and a free term $f(x,t)$ represented as

\begin{equation}
\overset{\symbol{94}}{L}U(x,t)+\overset{\symbol{94}}{N}U(x,t)=f(x,t)\text{,}
\tag{26}
\end{equation}%
where $\overset{\symbol{94}}{L}$ is the linear operator and $\overset{%
\symbol{94}}{N}$ is the nonlinear operator. According to the VIM, the ($n+1$)%
\underline{th} approximation solution of (26) can be given by the iteration
correction functional as [24, 25]

\begin{equation}
U_{n+1}(x,t)=U_{n}(x,t)+\int_{0}^{t}d\tau \lambda (\tau )[\overset{\symbol{94%
}}{L}U_{n}(x,\tau )+\overset{\symbol{94}}{N}\hat{U}_{n}(x,\tau )-f(x,\tau )]%
\text{, }n\geq 0\text{,}  \tag{27}
\end{equation}%
where $\lambda (\tau )$ is a Lagrangian multiplier and $\hat{U}_{n}(x,\tau )$
is considered as a restricted variation function, i. e; $\delta \hat{U}%
_{n}(x,\tau )=0$. Extreme the variation of the correction functional (27)
leads to the Lagrangian multiplier $\lambda (\tau )$. The initial iteration
can be used as the solution of the linear part of (26) or the initial value $%
U(x,0)$. As n tends to infinity, the iteration leads to the exact solution
of (26), i. e;

\begin{equation}
U(x,t)=\underset{n\rightarrow \infty }{\lim }U_{n}(x,t)\text{.}  \tag{28}
\end{equation}%
\qquad For linear problems, the exact solution can be given using this
method in only one step where its Lagrangian multiplier can be exactly
identified.

\section{Time-fractional KdV equation Solution}

The time-FKdV equation represented by (24) can be solved using the VIM by
the iteration correction functional (27) as follows:

Affecting from left by the fractional operator on (24) leads to

\begin{eqnarray}
\frac{\partial }{\partial \tau }\Phi (\xi ,\tau ) &=&~_{0}^{R}D_{\tau }^{\
\alpha -1}\Phi (\xi ,\tau )|_{\tau =0}\frac{\tau ^{\alpha -2}}{\Gamma
(\alpha -1)}  \notag \\
&&-\ _{0}^{R}D_{\tau }^{\ 1-\alpha }[A~\Phi (\xi ,\tau )\frac{\partial }{%
\partial \xi }\Phi (\xi ,\tau )+B~\frac{\partial ^{3}}{\partial \xi ^{3}}%
\Phi (\xi ,\tau )]\text{, }  \notag \\
0 &\leq &\alpha \leq 1\text{, }\tau \in \lbrack 0,T_{0}]\text{,}  \TCItag{29}
\end{eqnarray}

where the following fractional derivative property is used [18, 19]

\begin{eqnarray}
\ _{a}^{R}D_{b}^{\ \alpha }[\ _{a}^{R}D_{b}^{\ \beta }f(t)]
&=&~_{a}^{R}D_{b}^{\ \alpha +\beta }f(t)-\overset{k}{\underset{j=1}{\sum }}\
_{a}^{R}D_{b}^{\ \beta -j}f(t)|_{t=a}~\frac{(t-a)^{-\alpha -j}}{\Gamma
(1-\alpha -j)}\text{, }  \notag \\
k-1 &\leq &\beta <k\text{.}  \TCItag{30}
\end{eqnarray}

As $\alpha <1$, the Riesz fractional derivative $_{0}^{R}D_{\tau }^{\ \alpha
-1}$ is considered as Riesz fractional integral $_{0}^{R}I_{\tau }^{1-\alpha
}$ that is defined by [18, 19]

\begin{equation}
\ _{0}^{R}I_{\tau }^{\ \alpha }f(t)=\frac{1}{2}[_{0}I_{\tau }^{\ \alpha
}f(t)\ +~_{\tau }I_{b}^{\ \alpha }f(t)]=\frac{1}{2}\frac{1}{\Gamma (\alpha )}%
\int_{a}^{b}d\tau |t-\tau |^{\alpha -1}f(\tau )\text{, }\alpha >0\text{.} 
\tag{31}
\end{equation}%
where $_{0}I_{\tau }^{\ \alpha }f(t)$\ and $_{\tau }I_{b}^{\ \alpha }f(t)$
are the left and right Riemann-Liouvulle fractional integrals, respectively
[18, 19].

The iterative correction functional of equation (29) is given as

\begin{eqnarray}
\Phi _{n+1}(\xi ,\tau ) &=&\Phi _{n}(\xi ,\tau )+\int_{0}^{\tau }d\tau
^{\prime }\lambda (\tau ^{\prime })\{\frac{\partial }{\partial \tau ^{\prime
}}\Phi _{n}(\xi ,\tau ^{\prime })  \notag \\
&&-~_{0}^{R}I_{\tau ^{\prime }}^{1-\alpha }\Phi _{n}(\xi ,\tau ^{\prime
})|_{\tau ^{\prime }=0}\frac{\tau ^{\prime \alpha -2}}{\Gamma (\alpha -1)} 
\notag \\
&&+\ _{0}^{R}D_{\tau ^{\prime }}^{\ 1-\alpha }[A~\overset{\symbol{126}}{\Phi
_{n}}(\xi ,\tau ^{\prime })\frac{\partial }{\partial \xi }\overset{\symbol{%
126}}{\Phi _{n}}(\xi ,\tau ^{\prime })+B~\frac{\partial ^{3}}{\partial \xi
^{3}}\overset{\symbol{126}}{\Phi _{n}}(\xi ,\tau ^{\prime })]\}\text{,} 
\TCItag{32}
\end{eqnarray}%
where $n\geq 0$ and the function $\overset{\symbol{126}}{\Phi _{n}}(\xi
,\tau )$ is considered as a restricted variation function, i. e; $\delta 
\overset{\symbol{126}}{\Phi _{n}}(\xi ,\tau )=0$. The extreme of the
variation of (32) using the restricted variation function leads to

\begin{eqnarray*}
\delta \Phi _{n+1}(\xi ,\tau ) &=&\delta \Phi _{n}(\xi ,\tau
)+\int_{0}^{\tau }d\tau ^{\prime }\lambda (\tau ^{\prime })~\delta \frac{%
\partial }{\partial \tau ^{\prime }}\Phi _{n}(\xi ,\tau ^{\prime }) \\
&=&\delta \Phi _{n}(\xi ,\tau )+\lambda (\tau )~\delta \Phi _{n}(\xi ,\tau
)-\int_{0}^{\tau }d\tau ^{\prime }\frac{\partial }{\partial \tau ^{\prime }}%
\lambda (\tau ^{\prime })~\delta \Phi _{n}(\xi ,\tau ^{\prime })=0\text{.}
\end{eqnarray*}

This relation leads to the stationary conditions $1+\lambda (\tau )=0$ and $%
\frac{\partial }{\partial \tau ^{\prime }}\lambda (\tau ^{\prime })=0$,
which leads to the Lagrangian multiplier as $\lambda (\tau ^{\prime })=-1$.

Therefore, the correction functional (32) is given by the form

\begin{eqnarray}
\Phi _{n+1}(\xi ,\tau ) &=&\Phi _{n}(\xi ,\tau )-\int_{0}^{\tau }d\tau
^{\prime }\{\frac{\partial }{\partial \tau ^{\prime }}\Phi _{n}(\xi ,\tau
^{\prime })  \notag \\
&&-~_{0}^{R}I_{\tau ^{\prime }}^{1-\alpha }\Phi _{n}(\xi ,\tau ^{\prime
})|_{\tau ^{\prime }=0}\frac{\tau ^{\prime \alpha -2}}{\Gamma (\alpha -1)} 
\notag \\
&&+\ _{0}^{R}D_{\tau ^{\prime }}^{\ 1-\alpha }[A~\Phi _{n}(\xi ,\tau
^{\prime })\frac{\partial }{\partial \xi }\Phi _{n}(\xi ,\tau ^{\prime })+B~%
\frac{\partial ^{3}}{\partial \xi ^{3}}\Phi _{n}(\xi ,\tau ^{\prime })]\}%
\text{,}  \TCItag{33}
\end{eqnarray}%
where $n\geq 0$.

In Physics, if $\tau $ denotes the time-variable, the right
Riemann-Liouville fractional derivative is interpreted as a future state of
the process. For this reason, the right-derivative is usually neglected in
applications, when the present state of the process does not depend on the
results of the future development [3]. Therefore, the right-derivative is
used equal to zero in the following calculations.

The zero order correction of the solution can be taken as the initial value
of the state variable, which is taken in this case as

\begin{equation}
\Phi _{0}(\xi ,\tau )=\Phi (\xi ,0)=A_{0}\sec \text{h}^{2}(c\xi )\text{.} 
\tag{34}
\end{equation}

where $A_{0}$ and $c$ are constants.

Substituting this zero order approximation into (33) and using the
definition of the fractional derivative (25) lead to the first order
approximation as

\begin{eqnarray}
\Phi _{1}(\xi ,\tau ) &=&A_{0}\sec \text{h}^{2}(c\xi )  \notag \\
&&+2A_{0}c~\sinh (c\xi )~\sec \text{h}^{3}(c\xi )[4c^{2}B  \notag \\
&&+(A_{0}A-12c^{2}B)\sec h^{2}(c\xi )]\frac{\tau ^{\alpha }}{\Gamma (\alpha
+1)}\text{.}  \TCItag{35}
\end{eqnarray}

Substituting this equation into (33), using the definition (25) and the
Maple package lead to the second order approximation in the form

\begin{eqnarray}
\Phi _{2}(\xi ,\tau ) &=&A_{0}\sec \text{h}^{2}(c\xi )  \notag \\
&&+2A_{0}c~\sinh (c\xi )~\sec \text{h}^{3}(c\xi
)[4c^{2}B+(A_{0}A-12c^{2}B)\sec h^{2}(c\xi )]\frac{\tau ^{\alpha }}{\Gamma
(\alpha +1)}  \notag \\
&&  \TCItag{36}
\end{eqnarray}

The higher order approximations can be calculated using the Maple or the
Mathematica package to the appropriate order where the infinite
approximation leads to the exact solution.

\section{Results and calculations}

For small amplitude electron-acoustic waves, the time fractional Korteweg-de
Vries equation has been derived. The Riemann-Liouvulle fractional derivative
[18, 19] is used to describe the time fractional operator in the FKdV
equation. He's variational-iteration method [24, 25] is used to solve the
derived time-FKdV equation.

To make our result physically relevant, numerical studies have been made
using plasma parameters close to those values corresponding to the dayside
auroral zone [15, 34].

However, since one of our motivations was to study effects of time
fractional order ($\alpha $), the energetic population parameter ($\delta $)
and the hot to cold electron equilibrium densities ratio ($\gamma $) on the
existence of solitary waves. The electrostatic potential as a function of
space variable $\xi $ and time variable $\tau $ is represented in Figure (1)
that has a single rarefactive soliton shape. In Figure (2), the relation
between the hot to cold electron equilibrium densities ratio ($\gamma $) and
the amplitude of the electrostatic potential solitary wave $|\Phi (0,\tau )|$
is obtained at different values of the time variable $\tau $. It is seen
that the soliton amplitude decreases with the increase of ($\gamma $). In
addition the soliton amplitude $|\Phi (0,\tau )|$ against the fractional
order ($\alpha $) is represented in Figure (3) at different time variable
values. It is seen that $|\Phi (0,\tau )|$ increases with the increase of ($%
\alpha $). Figure (4) shows the relation between $|\Phi (0,\tau )|$ and the
energetic population parameter ($\delta $). The effect of the fractional
order ($\alpha $) on the electrostatic soliton amplitude $|\Phi (0,\tau )|$
for different values of the energetic population parameter ($\delta $) is
given in Figure (5). To compare our result (the amplitude of the
electrostatic potential $|\Phi (0,\tau )|$) with that observed in the
auroral zone, we choose a set of available parameters corresponding to the
dayside auroral zone where an electric field amplitude $E_{0}=100$ $\unit{mV}%
/\unit{m}$ has been observed [15, 34] with $T_{c}=5\unit{eV}$, $T_{h}=250%
\unit{eV}$, $n_{c0}=0.5\unit{m}^{-3}$ and $n_{h0}=2.5\unit{m}^{-3}$. These
parameters correspond to $\lambda _{Dh}\approx 7430\unit{cm}$ and the
normalized electrostatic wave potential amplitude $\Phi _{0}=\frac{%
E_{0}\lambda _{Dh}e}{k_{B}T_{h}}\approx 0.03\unit{V}$, which is obtained for
different values of ($\alpha $) and ($\delta $) [cf. Figure (5)]. This value
of the electrostatic potential amplitude is given at values of ($\alpha $)
and ($\delta $) as: ($\alpha =0.61$, $\delta =0.01$), ($\alpha =0.69$, $%
\delta =0.05$), ($\alpha =0.78$, $\delta =0.1$) and ($\alpha =0.9$, $\delta
=0.15$) at velocity $v=0.04$, $\gamma =5$ and time $\tau =10$.

In summery, it has been found that the amplitude of the electron acoustic
solitary waves as well as the parametric regime where the solitons can exist
are sensitive to the time fractional order.

We have stressed out that it is necessary to include the space fractional
parameter. This is beyond the scope of the present paper and it will be
include in a further work in electron-acoustic solitary wave. The
application of our model might be particularly interesting in the auroral
region.\pagebreak

\textbf{Figure Captions:}

Fig. (1): The electrostatic potential $\Phi (\xi ,\tau )$ against $\xi $ and 
$\tau $ at $\gamma =5$, $v=0.04$, $\alpha =0.5$ and $\delta =0.1$.

Fig. (2): The amplitude of the electrostatic potential $|\Phi (0,\tau )|$
against $\gamma $ at different values of time for $v=0.04$, $\alpha =0.5$
and $\delta =0.1$.

Fig. (3): The amplitude of the electrostatic potential $|\Phi (0,\tau )|$
against $\alpha $ at different values of time for $\gamma =5$, $v=0.04$ and $%
\delta =0.1$.

Fig. (4): The amplitude of the electrostatic potential $|\Phi (0,\tau )|$
against $\delta $ at different values of time for $\gamma =5$, $v=0.04$ and $%
\alpha =0.5$.

Fig. (5): The amplitude of the electrostatic potential $|\Phi (0,\tau )|$
against $\alpha $ at different values of $\delta $ for $v=0.04$, $\tau =10$
and $\gamma =5$.\pagebreak \pagebreak

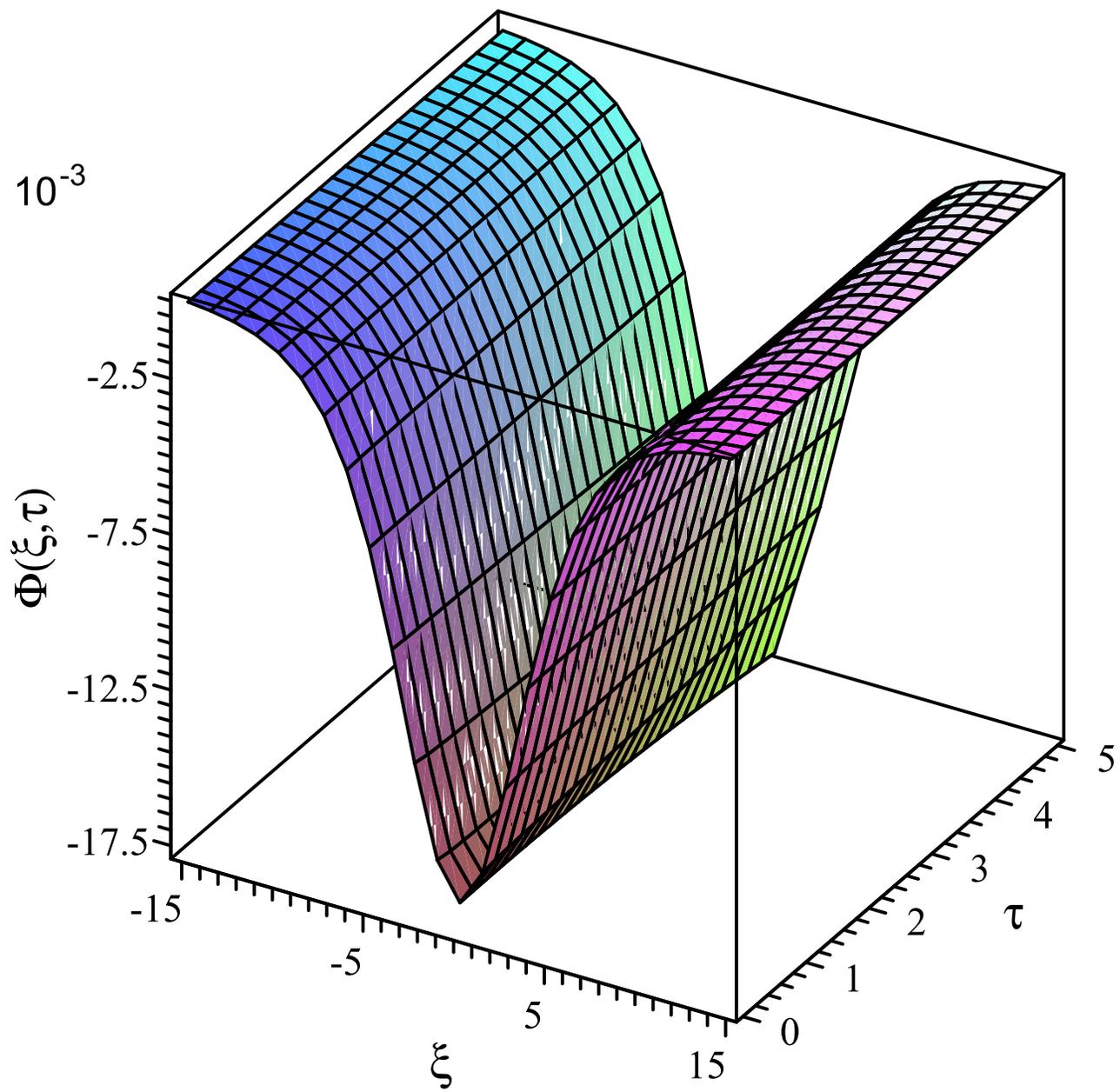
\end{document}